\documentclass[3p,times]{elsarticle}

\usepackage{ecrc}
\usepackage{color}

\volume{00}
\firstpage{1}
\journalname{Procedia Computer Science}

\runauth{Y. Nakamura et al.}

\jid{procs}
\jnltitlelogo{Procedia Computer Science}

\usepackage{amsmath,amssymb}
\usepackage{hyperref}

\def\cell{PowerXCell~8i}
\def\tpeak{t_\mathrm{peak}}
\def\texe{t_\mathrm{exe}}
\def\epsfp{\epsilon_\mathrm{FP}}

\begin{document}

\begin{frontmatter}

\dochead{International Conference on Computational Science, ICCS 2011}

\title{Lattice QCD Applications on QPACE}

\author[ts]{Y. Nakamura}
\author[re]{A. Nobile}
\author[de]{D. Pleiter}
\author[de]{H. Simma}
\author[re]{T. Streuer}
\author[re]{T. Wettig}
\author[ed]{F. Winter}

\address[ts]{Center for Computational Sciences, University of Tsukuba,
             Ibaraki 305-8577, Japan}
\address[re]{Department of Physics, University of Regensburg,
             93040 Regensburg, Germany}
\address[de]{Deutsches Elektronen-Synchrotron DESY,
             15738 Zeuthen, Germany}
\address[ed]{School of Physics, University of Edinburgh,
             Edinburgh EH9 3JZ, UK}

\begin{abstract}
  QPACE is a novel massively parallel architecture optimized for
  lattice QCD simulations.  A single QPACE node is based on the IBM
  {\cell} processor. The nodes are interconnected by a custom
  3-dimensional torus network implemented on an FPGA. The compute
  power of the processor is provided by 8 Synergistic Processing
  Units. Making efficient use of these accelerator cores in scientific
  applications is challenging. In this paper we describe our
  strategies for porting applications to the QPACE architecture and
  report on performance numbers.
\end{abstract}


\end{frontmatter}


\section{Introduction} 

Improving our understanding of the fundamental forces in nature by
means of numerical simulations has become an important approach in
elementary particle physics. In particular, computer simulations are
required to investigate the strong interactions because of their
non-perturbative nature. The theory which is supposed to describe the
strong interactions is Quantum Chromodynamics (QCD). The formulation
of the theory on a discrete space-time lattice is called lattice QCD
(LQCD) and has opened the path to numerical calculations.  Results
from such calculations are key input for the interpretation of data
obtained from experiments performed at existing and planned particle
accelerators.  The enormous cost of such experiments amply justifies
the numerical efforts undertaken in LQCD.

Since the availability of computing resources has been and continues
to be a limiting factor for making progress in this research field,
several projects have been carried out in the past aiming at the
development and deployment of massively parallel computers optimized
for LQCD applications.  Until the previous generation such special
purpose computers were typically based on a custom design including a
custom-designed processor, see, e.g., apeNEXT \cite{apenext} and QCDOC
\cite{qcdoc}. In QPACE, however, each node comprises a commodity
processor: An IBM {\cell}, one of the most powerful processors
available at project start.

From an architectural point of view QPACE may be considered to be a
cluster of (not particularly powerful) PowerPC cores with an attached
accelerator. Heterogeneous node architectures have recently become
more common among the most powerful supercomputers, which can be seen
from the top positions of the Top500 list, see Roadrunner (Los Alamos,
USA) or more recently Tianhe-1A (Tianjin, China).  Such more complex
hardware architectures pose significant challenges to the application
programmers who want to make efficient use of such powerful machines.

When porting applications to QPACE we applied two strategies. First,
highly optimized versions of kernels which dominate the overall
performance have been implemented aiming for a significant speed-up of
these kernels.  Second, the implementation of software interfaces has
been explored which allow one to use the accelerator cores while
completely hiding the architectural details from the application
programmer when implementing the remaining part of the code. This
allows one to increase the fraction of code which is accelerated.

The performance can furthermore be improved by choosing an algorithm
which is optimal for a given architecture.  However, on heterogeneous
architectures it is often much more difficult to assess the interplay
between machine performance and algorithmic performance.  The machine
performance can be defined as a set of relevant performance numbers
which can be measured during execution of a computational task in
terms of a hardware related metric, e.g., number of floating-point
operations per time unit executed by a particular set of hardware
units.  The algorithmic performance refers to the number of atomic
sub-tasks, e.g., matrix-vector multiplications, to solve a particular
problem, e.g., to compute the solution of a system of linear
equations.

After an overview of the QPACE architecture we will discuss in
section~\ref{section:application} the requirements of the application
for which this machine has been designed. In sections
\ref{section:swintf} and \ref{section:kernel} we will discuss two
independent approaches to the porting of application codes to QPACE.
In sections \ref{section:chroma} and \ref{section:bqcd} we report
on performance results.

\section{QPACE architecture}\label{section:architecture} 

\begin{figure}[ht]
\begin{center}
\includegraphics[height=50mm]{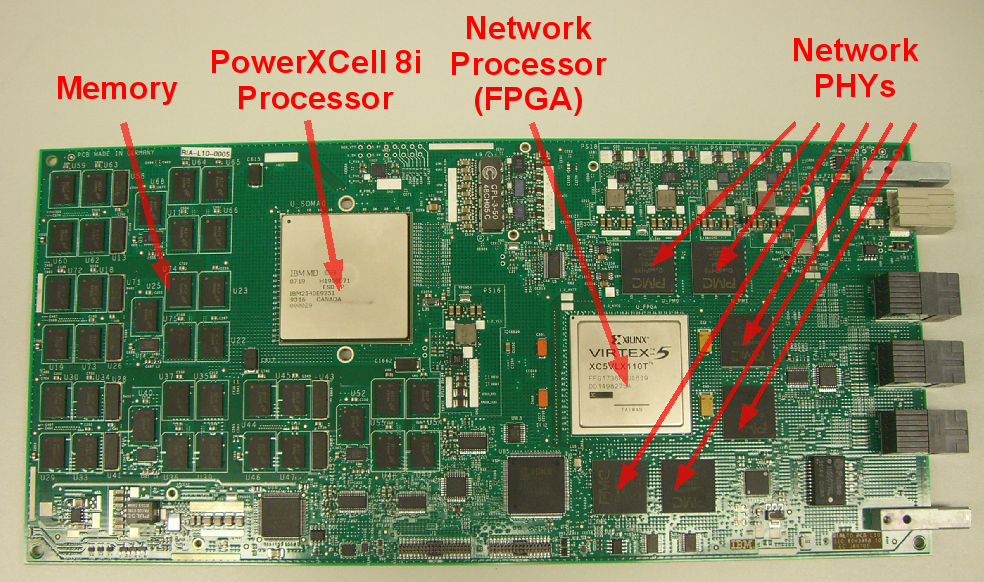}\hfill
\includegraphics[height=50mm]{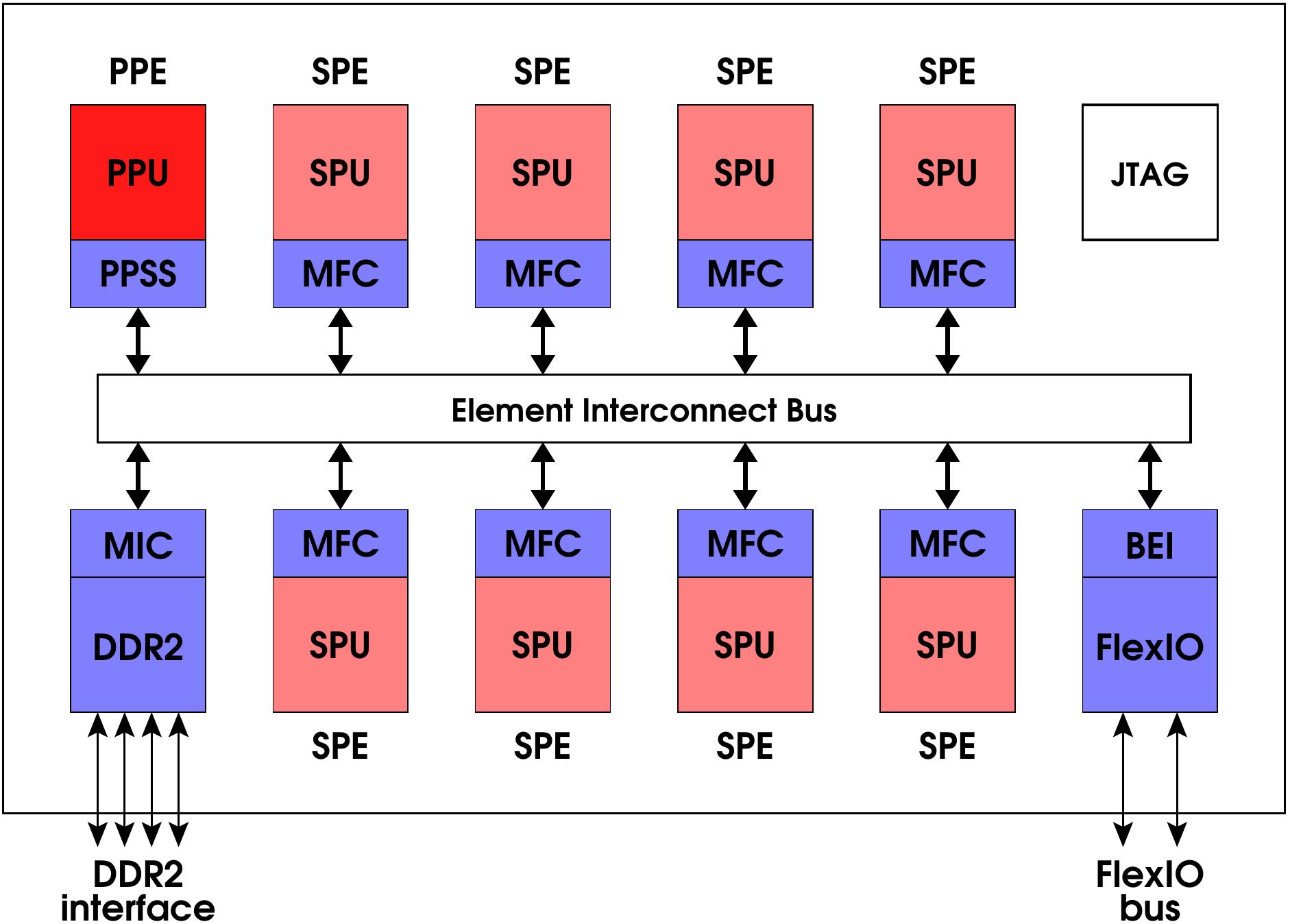}
\end{center}
\vspace*{-3mm}
\caption{\label{fig:rialto}QPACE node-card (left) and simplified block
  diagram of the {\cell} processor (right).}
\end{figure}

The main building block of the QPACE architecture is the node card,
see Fig.~\ref{fig:rialto} (left). The commodity part of the node
consists of a {\cell} processor and 4~GBytes of main memory. To
realize a custom network a Field Programmable Gate Array (FPGA) and
six physical transceivers (PHY) have been added. The FPGA implements
the Network Processor (NWP), a custom I/O fabric which is directly
connected to the processor and which furthermore includes 6 links
through which the node is connected to its nearest neighbors within a
3-dimensional torus.  The PHYs implement the physical layer of the
torus network links.

The {\cell} processor is an enhanced version of the Cell processor
which has been developed by Sony, Toshiba, and IBM, with Sony's
PlayStation 3 as the first major commercial application.  The {\cell}
is the second implementation of the Cell Broadband Engine
Architecture~\cite{cbea}, which unfortunately has been discontinued.
The most important enhancements are the support for high-performance
double precision operations and IEEE-compliant rounding.  The
processor comprises multiple cores, see Fig.~\ref{fig:rialto} (right),
including the Power Processing Element (PPE), which is a standard
PowerPC core. On this core Linux is running and the execution of
applications is started. To make full use of the processor's
compute power the application has to start threads on the 8
Synergistic Processing Elements (SPE).  All 9 cores, the memory
interface as well as the I/O interface are interconnected via a
ring-bus. This bus has a very high bandwidth of up to 200 GBytes/sec.

The SPEs have a non-standard architecture, which, however, is very
suitable for LQCD applications. Each of them consists of a Synergistic
Processing Unit (SPU) and a Memory Flow Controller (MFC). In each
clock cycle the SPU can perform a multiply-add operation on a vector
of 4 (2) single (double) precision floating-point numbers.  With a
clock frequency of 3.2 GHz this gives a peak floating-point
performance per SPU of 25.6 or 12.8~GFlops in single or double
precision, respectively.  The memory hierarchy of the processor is
non-trivial. Each of the SPUs has its own Local Store (LS), a 256
kBytes on-chip memory.
Loading and storing data from and to other devices attached to the
ring-bus is handled by the Direct Memory Access (DMA) engine in the
MFC. The interface to the external memory provides a bandwidth of
25.6~GBytes/sec. This interface is shared by all SPEs, thus reducing
the bandwidth per clock cycle and SPE to 1 Byte peak.

Thanks to an innovative water-cooling system a QPACE rack can host up
to 256 nodes, i.e., the aggregate peak performance per rack is 52 or
26 TFlops in single or double precision, respectively.  For further
details of the QPACE hardware see \cite{Baier:2009yq,qpace-enahpc}.

The custom torus network \cite{Pivanti:2011hd} has been optimized for LQCD
applications.  Since the communication patterns in these applications
are typically homogeneous and symmetric a two-sided communication
model has been adopted.  When node $A$ initiates transmission of a
message to node $B$ it has to rely on node $B$ initiating a
corresponding receive operation.  This communication model minimizes
the protocol overhead since it avoids handshake between transmitter
and receiver.  However, it leaves the responsibility to the programmer
to keep the order of the communication operations and the
parameters (e.g., message size) on transmitter
and receiver side consistent.

When a processor wants to inject data into the network it has to write
the data to the corresponding link in the NWP. The typical use case is
that the transmit buffer resides in the LS and the DMA engine of the
attached MFC is used for moving the data to the FPGA. On the receiving
side the processor has to provide a credit to the NWP which defines
how many data packets can be written to the receive buffers. These
buffers can either be located on-chip (i.e., in LS) or off-chip (i.e.,
in main memory). The packets have a fixed length and consist of a
4~Byte header, a 128~Byte payload, and a 4~Byte checksum. When a
packet arrives at the destination link and a matching credit is
available a DMA engine in the NWP will write the payload into the
receive buffer. Once a credit has been consumed a notification is
generated and the transfer is completed.  The network supports virtual
channels to allow different pairs of cores to use the same physical
link.

\section{Application requirements}\label{section:application} 

Simulations in lattice QCD can be split in multiple stages:
\begin{enumerate}
\item Generation of gauge field configurations
\item Measurement of physical observables on these gauge field configurations,
      e.g., computation of correlation functions
\item Analysis of the measurements
\end{enumerate}
The first two stages are by far the most compute intensive. For
instance, the generation of a typical ensemble of gauge field
configurations may require dozens of sustained TFlops years.
We have ported two application programs which can be used for these first
stages of simulation to the Cell processor or QPACE:
\begin{itemize}
\item BQCD \cite{Nakamura:2010qh} is a program for generating gauge
      configurations with $2$ or $2+1$ flavors of different types of
      Wilson fermions. It is extensively
      used on various massively parallel machines, including
      machines with x86-based nodes, IBM BlueGene, and QPACE.
\item Chroma \cite{Edwards:2004sx} is a LQCD software system which consists of
      several applications including a program to compute
      various physical observables.
      It is implemented on top of the QCD Data Parallel (QDP++) library.
\end{itemize}

In both applications a large fraction of the execution time is spent on
solving a system of linear equations of type
\begin{equation}
M\,\phi = b \quad \mathrm{or} \quad M^\dagger M\,\phi = b \label{eq:linear}\,.
\end{equation}
The exact form of the matrix $M$ may vary in different calculations. This
reflects the fact that there is no unique way of formulating the
physical theory, which is defined in a space-time continuum, on a discrete
lattice. In the following we will only consider the case of Wilson or
Clover fermions, where $M$ is not Hermitian.
All formulations have in common that $M$ is a huge
but sparse complex matrix. Therefore, iterative algorithms are used to
compute a solution for the system of linear equations. Conjugate Gradient
is probably the best known (although not necessarily the most efficient)
algorithm used for this kind of problems.

Any of the iterative algorithms can be split into a sequence of computational
tasks which involves matrix-vector multiplications.
From a computational point of view the latter is the most demanding step
since it is compute intensive and involves communication of data between the
nodes. On most architectures the performance of this task is limited by
the amount of data which can be transferred within the node and between the
nodes. It is therefore instructive to consider the information flow 
function $I_{x,y}^k(N)$ which defines the amount of data that have
to be transferred between two storage devices $x$ and $y$ (e.g., main memory and
register file) for a certain subtask $k$ and subtask size $N$.
The link or processing device which connects the storage devices can be
characterized by the bandwidth $\beta_{x,y}$ and start-up
latency $\lambda_{x,y}$. An estimate of the time $t_k$ needed to
execute the subtask $k$ is given by
\begin{equation}
t_k(N) = \lambda_{x,y} + \frac{I_{x,y}^k(N)}{\beta_{x,y}}\,.
\end{equation}
If we assume that in first approximation several subtasks, e.g., transfer
between main memory and processor as well as floating-point computations,
can run concurrently, an optimistic estimate of the execution time is
obtained taking
\begin{equation}
\texe \simeq \max_k t_k\,.
\end{equation}
Let us denote by $\tpeak$ the minimal compute time for the
floating-point operations of a task that could be achieved with an
ideal implementation.  The floating point efficiency $\epsfp$ for a
given task is then defined as $\epsfp = \texe / \tpeak$.

As an example let us consider the operation
\begin{equation}
U_{x,a,b} \leftarrow \sum_c V_{x,a,c}\,W_{x,c,b}\,,\label{eq:su3vecmul}
\end{equation}
where $U$, $V$, $W$ are arrays (of length $N$) of double precision
complex $3\times 3$ matrices, and $x = 0,\ldots,N-1$.
These SU(3) matrices occur at different places in LQCD applications.
We will assume
$N$ to be large such that the arrays cannot be kept in the LS of the
{\cell} processor and have to be stored in main memory (MM).  The
register file is, however, large enough to guarantee full data re-use
during the matrix multiplications.  Therefore, for each $x$ we have to
load $18\cdot16$ and store $9\cdot16$~Bytes, i.e.,
$I_\mathrm{MM,LS}(N) = N\,432\,\mathrm{Bytes}$.  For the ideal case of
full memory bandwidth saturation (and negligible or hidden latency)
we find $t_\mathrm{MM,LS}(N) = N\,432\,\mathrm{Bytes} /
(25.6\,\mathrm{GBytes/s})=N\cdot17$ ns.  For each $x$ we furthermore
have to perform 108 fused vector multiply-adds.  Ignoring data
dependencies and shuffle operations the time for this subtask is
$t_\mathrm{FP} = N\,108 / 3.2\,\mathrm{GHz} /
N_\mathrm{core}=N\cdot4.2$ ns. The sustained performance of this
operation is thus dominated by the memory bandwidth.

\section{Optimized software interfaces}\label{section:swintf} 

In LQCD applications a large fraction of the computational effort is
spent in a few kernel routines. But once the accelerator cores provide
a significant speed-up of these routines, the remaining part starts to
dominate the overall execution time. This is a well-known result of
Amdahl's law which tells us that the total speed-up $S_\mathrm{total}$
is limited by the fraction $P$ which can be accelerated by a factor
$S$:
\begin{equation}
S_\mathrm{total} = \frac{1}{(1-P) + P/S}\,.
\end{equation}
Let us assume that $P$ is the fraction spent for solving the system of
linear equations described in the previous section. In this case $P$
strongly depends on the simulation parameters but is in practice
$\lesssim 90\,\%$.

If $S$ is large it may become more efficient (or, in fact, necessary)
to increase $P$ in order to improve $S_\mathrm{total}$. Our strategy
to increase $P$ was to port a software interface to the {\cell}
processor to hide the hardware details from the application
programmer. This strategy has been extensively used for the Chroma
application suite which is implemented on top of QDP++, which provides
a data-parallel programming environment suitable for implementing LQCD
applications.  Porting Chroma to another architecture basically
reduces to a port of QDP++.

QDP++ is a C++ library which can be considered to be an active
library, i.e., a library which takes an active role in generating code
and interacts with programming tools \cite{veldhuizen}. This leads to
particular challenges when porting this code since, e.g., the list of
instantiated functions will only be known after compilation.

\begin{figure}[ht]
\vspace*{-1mm}
\begin{center}
\includegraphics[height=62mm]{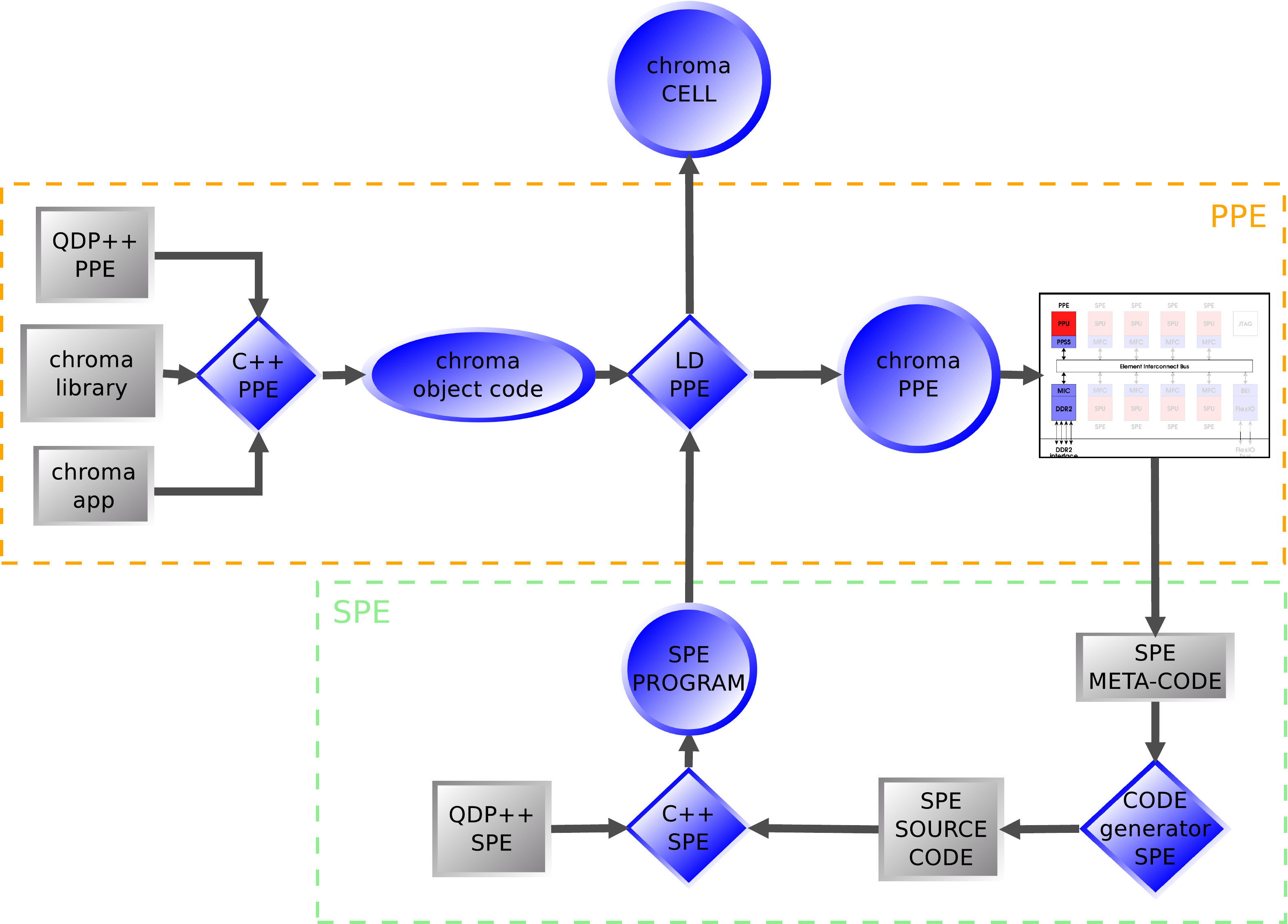}
\end{center}
\vspace*{-3mm}
\caption{\label{fig:compile}
The build process of Chroma for the {\cell} processor:
Starting from the left first an executable for the PPE
is generated. During a test execution of this executable a list
of instantiated functions is retrieved. This is used by
a code generator which generates SPE code.
This code and a library is used to produce a new executable.}
\end{figure}

To port this library to the {\cell} processor the following approach has
been chosen \cite{winter-phd}:
\begin{enumerate}
\item Identification of all QDP++ functions involved in a particular run
      of the main application by performing a test run.
\item Automated generation of SPE code for each of the individual
      functions.
\item Re-generation of the main executable now including the SPE code
      and call-outs to the SPE threads from the PPE.
\end{enumerate}
The resulting application build process is shown in Fig.~\ref{fig:compile}.

The list of instantiated functions is obtained during the test run,
i.e., the PPE-only version of the executable is run and so-called
pretty-functions are streamed into the meta-code database.  A
pretty-function is a C-style string variable available at run time
containing the function's name, return type, and arguments.  The meta
code collected during this test run is processed by an SPE code
generator which constructs for each of the pretty-functions a C++
function restoring the original mathematical semantics but this time
with additional support for the accelerator architecture.  The
mathematical function is implemented using the lattice-wide data types
and operations provided by QDP++.  Support for the accelerator
architecture was integrated by implementing a light-weight version of
QDP++ for this type of processor cores.  In order to keep the size of
all instructions which will be loaded into the LS at the same time
small it is mandatory to use the code overlay feature provided by the
IBM Cell development tools.

Most functions operate on vectors such that the elements are processed
independently of each other, see the example in
Eq.~\eqref{eq:su3vecmul}.  To parallelize such an operation on
multiple cores the vector is split into disjoint parts and no further
dependencies need to be taken into account.

Since most of the operations are limited in performance by the time
needed to transfer data from MM to LS and back using explicit DMA
operations, special care is needed to optimize the bandwidth usage and
to hide latencies. The latter is achieved by employing multi-buffering
techniques. The required memory in the LS is managed by a custom
pool-based memory allocator. The software needs to determine the
transfer sizes, which depend on the size of the input/output vector
elements, the pool size, and memory address alignment requirements.
The pool size is fixed before compile time while the size of the
vector elements has to be determined at compile time. The vector
length typically is a run-time parameter.

In QDP++ data types are constructed in a nested manner leveraging
template programming techniques. It is therefore possible to manually
override the generic construction of types at any level of type
construction by means of class template specializations. This
technique is exploited in particular to improve the floating-point
performance, e.g., by transforming scalar operations into vector
operations which make use of the full width of the floating-point unit
in the SPE (and as a by-product eliminate shuffle operations).

The strategy outlined above has been successfully implemented and tested
on a single {\cell} processor. Performance results are reported in
section~\ref{section:chroma}. The work to extend this to a parallelized
version on QPACE is ongoing.

\section{Highly optimized application kernel}\label{section:kernel} 

In the following we consider two algorithms for solving the equations
given in \ref{eq:linear}. 
One is the Conjugate Gradient (CG), which exploits the symmetry and positive
definiteness of the normal operator $M^\dagger M$, and the other is
based on a domain decomposition strategy.

CG gained its popularity due to good convergence properties,
simplicity, and robustness. The most intensive numerical task in the
CG algorithm is the matrix-vector product.
Almost all of the optimization effort is usually spent on this particular task.
In the case of Wilson-like fermions assuming an optimal implementation
of the multiplication by the matrix $M_W$
(i.e., maximum data reuse and minimum traffic on the memory bus) it
was found that the performance is bounded by the main memory bandwidth
$\beta_\text{MM}$ to $\epsfp=34\%$ assuming the peak value for
$\beta_\text{MM}$ \cite{Belletti:2007pp}. For a real implementation
$\epsfp=24\%$ has been measured \cite{andrea-phd}, which is in good
agreement with the upper limit given by this simple model.
The considered implementation is subject to
a lattice-size constraint in order to achieve maximum data reuse.
The maximum local lattice size for a single-
(double-) precision computation is limited to $L_0\times 10^3$
($L_0\times 8^3$), where $L_0$, the temporal extent of the lattice, is
not constrained. This limitation comes from the size of the local
store.  This optimal implementation is also able to tolerate a network
latency of a few microseconds, i.e., $O(10,000)$ clock cycles.

It is, however, common to use CG to solve the Schur complement
equation arising from the even-odd decomposition of the matrix $M_W$
operator.  In this case both memory and network accesses become more
problematic.  During the application of the operator, the vectors
on the borders of the local lattice must be communicated twice, since
this operator couples next-to-nearest-neighbor lattice points. A
practical implementation requires two sweeps of the
lattice, each of which requires all of the so-called link variables
to be loaded
while doing half of the needed floating-point operations.  The memory
bandwidth limitation thus becomes more critical and the estimated
performance on a single node, taking into account the efficiency of the memory,
goes below $20\%$.

In order to circumvent these problems, including the restrictions on the
choice of the local lattice sizes, we
consider the SAP-FGCR algorithm proposed by L\"uscher
\cite{Luscher:2003qa}, who also demonstrated the algorithmic
speed-up in realistic use-cases.
The Schwarz Alternating Procedure (SAP)
belongs to the class of domain decomposition methods. The lattice is
divided into non-overlapping blocks which are
checkerboard-colored. One iteration (cycle) of the SAP proceeds as
follows: first the Wilson-Dirac equation is solved on all the
\emph{white} blocks, then the residue is updated both on the
\emph{white} blocks and on the internal border of the \emph{black}
blocks, then the equation is solved on the \emph{black} blocks, with
the source being the updated residue, and finally the residue is
updated on the \emph{black} blocks and on the internal border of the
\emph{white} blocks.  The blocks are solved with a fixed number of
iterations using a simple iterative solver (MR) and Dirichlet boundary
conditions.  In this procedure the only step which requires network
communications is the update of the residue on neighboring blocks. (We
assume that individual blocks do not extend beyond a single
processor.) Since the updated residue is needed only after half of the
blocks are solved, it is possible to completely overlap communication
and computation. The procedure is thus able to tolerate very high
network latencies without performance loss. The network bandwidth
requirement is a fraction $1/(n_\text{it}+1)$ of the requirement of
the equivalent Wilson-Dirac operator on the whole lattice, where
$n_\text{it}$ is the number of iterations used in the block solver.

The SAP procedure is used as a preconditioner inside an FGCR
iteration. The outer FGCR iteration is able to tolerate variations in
the preconditioner since the preconditioner is obtained with an
iterative method and thus is not a stationary operator. FGCR is
mathematically equivalent to FGMRES \cite{saad} and requires one to
store two vectors per iteration. This requirement translates in the
practical need to restart the FGCR recursion. Mixed precision is
implemented through iterative refinement \cite{itref}. All operations
can proceed in single precision, and only when the FGCR recursion is
restarted, the residue is computed in double precision and the
double-precision solution is updated by adding the single-precision
correction.  The most time-consuming task in the SAP is the solution
of the Wilson-Dirac equation on the blocks. Since the block size can
be chosen such that all the data necessary for a block solve fit in
the local store, SAP is able to achieve a high sustained performance
by making efficient use of the local store.  The local lattice
size is only constrained to be a multiple of the block size.

In order to allow for the maximum possible block size and to have the
flexibility to scale the algorithm to a large number of nodes, we
choose to parallelize the block solver over the 8 SPEs.  The block is
divided along the temporal direction across the 8 SPEs.  This
constrains the block size in this direction to be a multiple of 8,
which in practice is not a severe restriction.
Our implementation overlaps communication and
computation also for on-chip parallelization.

All the floating-point intensive code was written using vector
intrinsics, and different data layouts were analyzed theoretically
before implementing the code. The data layout has a strong impact on the
performance, and trade-offs may be necessary to simplify programming.
Lattice points are ordered such that points belonging to
the same block are contiguous. Inside the blocks, lattice points are
divided into two sets, even and odd. In this way DMA operations
performed to move block fields between main memory and local store are
greatly simplified and the performance is maximized.  For the spinor
layout, different possibilities were analyzed by counting the
floating-point and shuffle instructions involved in the application of
the Wilson-Dirac operator for the different layouts. The final choice
is such that the indices, from the slowest to the fastest, are
\emph{color, spinor, complex}.
The SU(3) matrices, consisting of 72-Byte arrays in
single precision, do not satisfy the basic 16-Byte alignment
constraint coming from the size of the registers.  We have thus chosen
to pad the corresponding data structures to 80 Bytes. While this choice introduces
an overhead by wasting a fraction of the memory bandwidth, it greatly
simplifies the code by avoiding the manual alignment operations that
would otherwise be necessary.

Complete overlap of memory accesses with computation can in principle
be achieved for sufficiently large on-chip memory.  We, however, have chosen to
use the maximum possible block size in order to maximize the block
solver performance.  This limits the amount of data that can be
prefetched.

Denoting the SAP operator by $M_\text{sap}$, a simple but effective
estimation of its performance is given by
\begin{equation}
  \epsilon_\text{Msap}=\frac{T_\text{bs-peak}\times(n_\text{it}+1)/n_\text{it}}{\max[T_\text{bs}\times(n_\text{it}+1)/n_\text{it} +  T_\text{LM}/\epsilon_\text{LM}\,,\,  T_\text{tnw}]}\,,\label{eq:sapmodel}
\end{equation}
where $T_\text{bs-peak}$ is the time spent in $n_\text{it}$ iterations
of the block solver assuming peak floating-point performance, the
factor $(n_\text{it}+1)/(n_\text{it})$ takes into account the
operations needed for the even-odd preconditioning on the blocks,
$T_\text{bs}$ is the actual time spent in the block solver for
$n_\text{it}$ iterations, $T_\text{LM}$ is the time spent moving data
between main memory and local store assuming peak main-memory
bandwidth, $\epsilon_\text{LM}$ is a coefficient that models the
efficiency of the memory subsystem, and $T_\text{tnw}$ is the time
needed for network communication.  It is assumed here that there is no
overlap between computation and main-memory accesses while there is
complete overlap between computation and network communication.  The
coefficient $\epsilon_\text{LM}$ was determined in micro-benchmarks to
be about 0.8.  

\begin{table}[h]
  \centering
  \begin{tabular}{lccccccccc}
    block size   &$\epsilon_\text{bs}$ (m) & $\epsilon_\text{Msap}$ (e) & $\epsilon_\text{Msap}$ (m) & $T_\text{FP}$  & $T_\text{LM}$ &$T_\text{tnw}$(1.5)& $T_\text{tnw}$(3.0) \\
    \hline \\[-3mm]
    8$\times$4$\times$8$\times$4 &    $36\%$                    & $25.8\%$ & $25.9\%$ &  458     & 145 & 128 & 64 \\
    8$\times$2$\times$6$\times$6&    $34\%$                    & $24.1\%$ & $23\%$ &  258    & 89 & 128 & 64 \\
    8$\times$2$\times$2$\times$10&    $30\%$                    & $21.7\%$ & $19.3\%$ &  172       & 52  & 94 & 47 \\
    \end{tabular}
    \caption{Performance numbers for the SAP preconditioner as explained
      in the text.} 
  \label{table}
\end{table}
Table~\ref{table} shows the performance of the block solver,
$\epsilon_\text{bs}$, and of the whole preconditioner,
$\epsilon_\text{Msap}$, for different block sizes.  The measured
performance (m) agrees very well with the estimated one (e), which is
obtained from expression \eqref{eq:sapmodel} assuming 4 block-solver
iterations.  $T_\text{FP}$, $T_\text{LM}$, $T_\text{tnw}$(1.5), and
$T_\text{tnw}$(3.0) are, respectively, the time spent doing
floating-point operations, moving data between local store and main
memory, and performing network communications assuming 1.5 GB/s and
3.0 GB/s bandwidth. Times are expressed in thousands of clock cycles
per block.

The structure of the SAP algorithm fits nicely the features of the
network.  After one block is solved, the data necessary for the update
of the residue on neighboring blocks residing on remote nodes are
available and sent via a DMA put to the remote nodes directly from the
SPEs. Communications are either between two cores of the same node
or between two nodes. This difference is, however, hidden in the SPE
code because it is controlled by the addresses provided by the
thread running on the PPE.
All the complexity associated with the network is thus handled by
the PPE control code which is also responsible for issuing network
receive commands. This simplifies coding significantly.

\section{Application performance: QDP++/Chroma}\label{section:chroma} 

\begin{figure}[ht]
\begin{center}
  \includegraphics[scale=0.57]{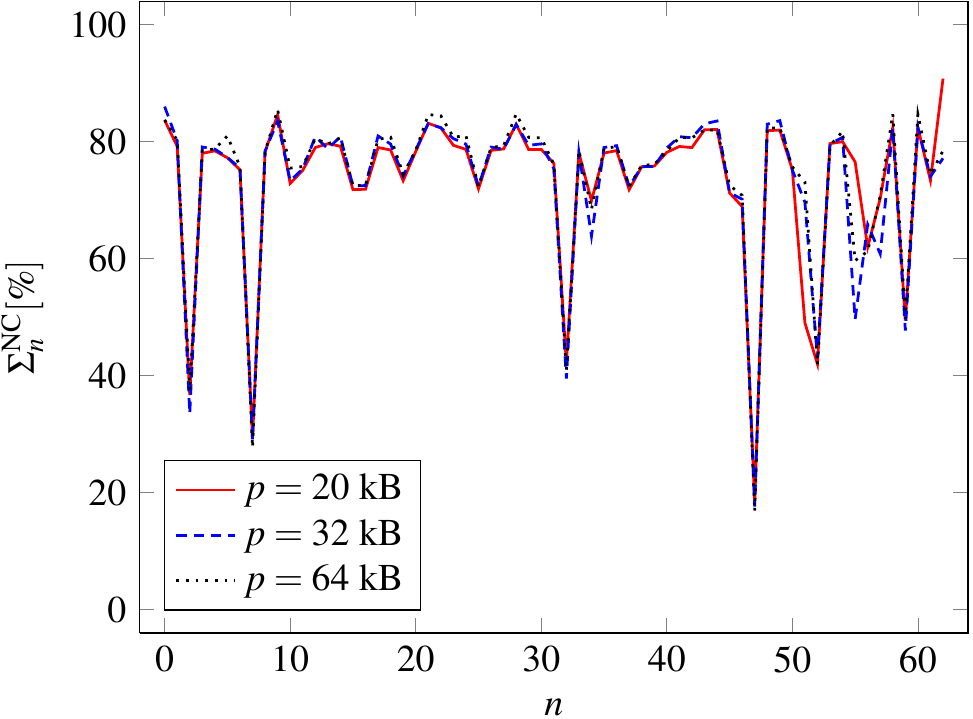}\hspace*{20mm}
  \includegraphics[scale=0.57]{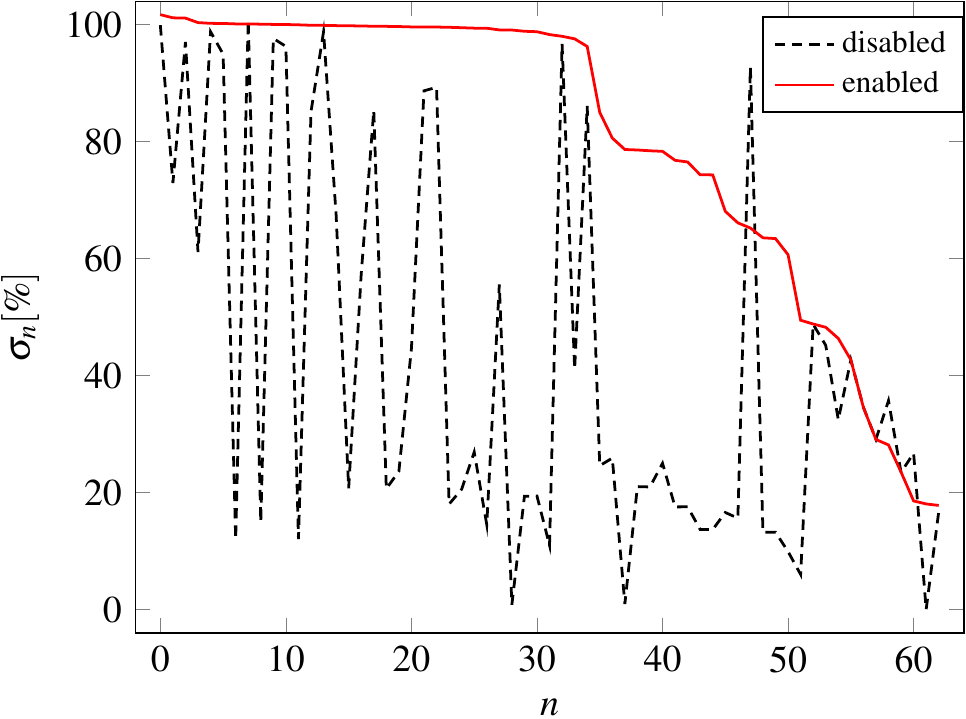}
  \vspace*{-6mm}
\end{center}
\caption{\label{fig:qdp}%
  The left pane shows the memory bandwidth saturation for all
  investigated QDP++ functions $f_n$ with SPE computations being
  switched off.  The different lines correspond to results for
  different pool sizes $p$. The right pane shows the relative change
  in memory bandwidth saturation $\sigma_n$ with and without template
  specializations (values $\sigma_n > 100\%$ are due to small fluctuations
  in the execution times).}
\end{figure}

We start our investigation of the performance of our QDP++ port by
considering the case of no computation, i.e., only memory transfer
operations are performed.  This is sensible because we expect the
performance of a large portion of the functions to be limited by the
bandwidth between MM and LS.  Fig.~\ref{fig:qdp} (left pane) shows, for
a large set of QDP++ functions $f_n$, the memory bandwidth saturation
$\Sigma^\mathrm{NC}_n$, where each of the functions has been assigned
an index $n$. The explicit mathematical form of the individual
functions cannot easily be recovered due to their construction via the
expression template technique.  We observe that for most of the
functions a good memory bandwidth saturation of about 80\% is
achieved. Only for a few functions the memory bandwidth is small. It
turned out that in these particular cases the load or store operations
could not be organized in an efficient way, e.g., because only a few
Bytes per vector element have to be transferred. However, the impact
on the overall performance of functions with only a few Bytes to
transfer is small.  The pool size $p$ seems to have little impact on
the memory bandwidth saturation for most of the functions.

Next we consider the relative change of the memory bandwidth saturation
when computations are switched on:
\begin{equation}
\sigma_n = \frac{\Sigma_n}{\Sigma^\mathrm{NC}_n}\,.
\end{equation}
In the right pane of Fig.~\ref{fig:qdp} $\sigma_n$ is shown for two
cases, using either generic code or template specializations.  At this
stage we have only executed an optimized version of the complex
multiplication.  The benchmark result shows the impact of using this
optimized version instead of the scalar version of complex
multiplication, which requires many more shuffle, load, and store
operations.  We also observe that more than half of the functions do
not require further optimization in terms of the number of
floating-point operations since they already execute with $\sigma_n
\simeq 100\%$.

However, for the other functions $\sigma$ can be increased by
supplying further optimized code via template specializations.  For
instance, the function $n=36$ corresponds to the operation defined in
Eq.~\eqref{eq:su3vecmul}, which is memory-bandwidth limited.  Here the
relative memory-bandwidth saturation is reduced by 20\% when
computation is enabled due to generic code for the matrix
multiplication. We expect that for an optimized implementation of the
matrix multiplications $\sigma_{36} \simeq 100\%$ could be achieved.

To test the performance of our QDP++ port in a full LQCD application
we measure the execution time for a particular application in Chroma
(which computes the hadron spectrum). This part does not involve calls
to optimized kernel routines and would thus usually not make use of
the accelerator cores.  On a dual-core 2.0~GHz Intel Xeon processor
5130 we measured an execution time of 83.5~sec.  On an IBM QS22 blade
using a single 3.2~GHz {\cell} processor we measured 617.9~sec for a
PPE-only version of Chroma.  Using our light-weight version of QDP++
for the accelerator cores, the execution time on the same system goes
down to 142.4~sec.  Thus, although the performance on the {\cell}
processor is significantly smaller compared to the Xeon processor, a
speed-up of over 4x is obtained from the PPE-only to the accelerated
version. In practice this speed-up is sufficient because only a
relatively small fraction of the compute time is spent on this part of
the program.

\section{Application performance: BQCD}\label{section:bqcd} 

BQCD implements the Hybrid Monte Carlo algorithm for generating
gauge field configurations.
The fraction of time spent for solving the linear system of equations
strongly depends on the simulation parameters.
For the most expensive runs currently performed by the QCDSF collaboration
(i.e., runs with $m_\pi \simeq 170\,\mbox{MeV}$ and a spatial
lattice extent $L_S \simeq 3.6\,\mbox{fm}$)
the BlueGene/P and QPACE versions of BQCD spend
about 95\% and 75\% of the total execution time
in the solver, respectively.
On QPACE we use the SAP-based solvers described in section~\ref{section:kernel}.
These solvers have been ported to the SPEs where they reach a performance of
$\epsfp\approx 20\%$ of the single-precision peak. Also some of the other
most time-consuming kernels of BQCD have been ported to the SPEs. However,
the rest of the code runs on the PPE which explains (at least partially) why a significantly larger
fraction of time is spent in this part compared to the BlueGene/P version.

\begin{figure}[ht]
  \begin{center}
    \vspace*{-1mm}
    \includegraphics[scale=0.57]{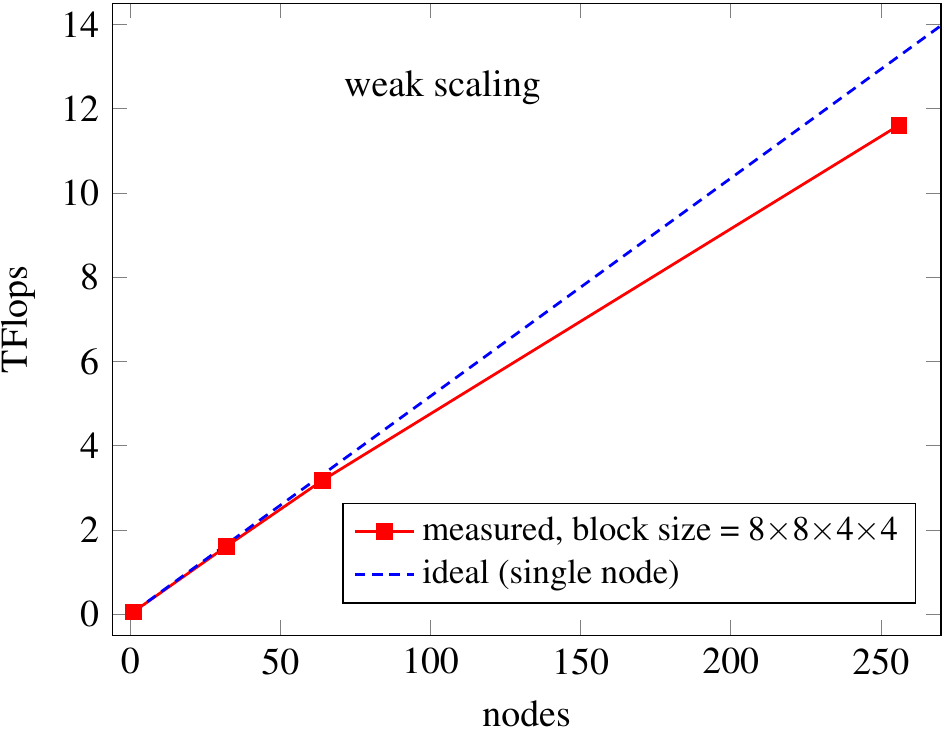}\hspace*{20mm}
    \includegraphics[scale=0.57]{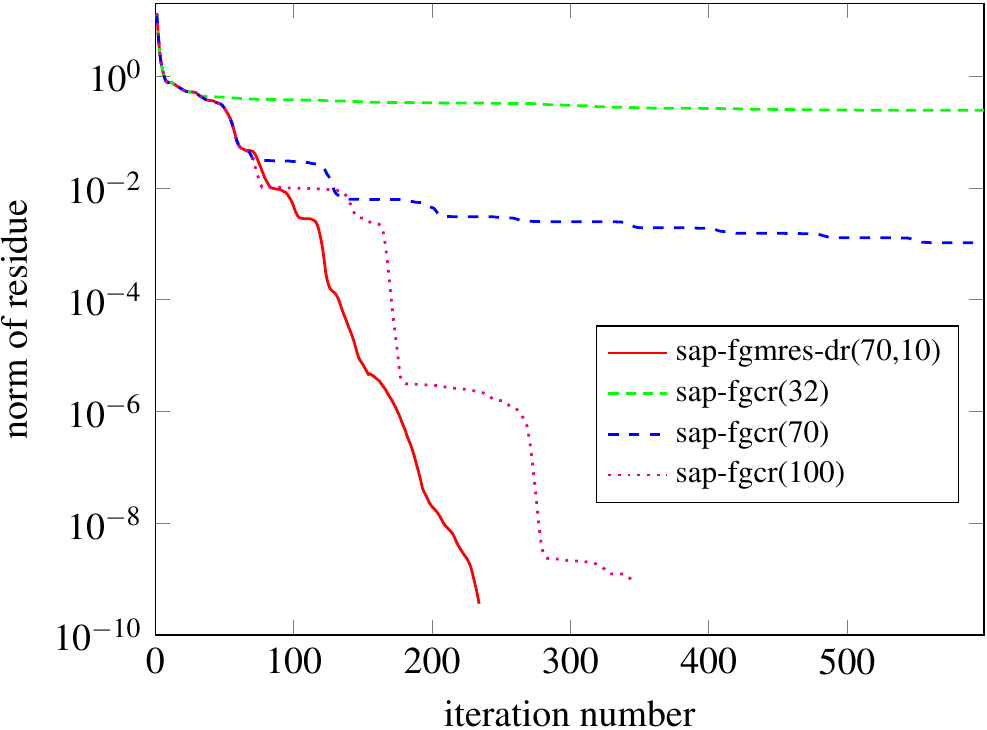}
    \vspace*{-6mm}
  \end{center}
  \caption{\label{fig:sapdr}%
    The left pane shows the scaling of the SAP preconditioner performance
    as a function of the
    number of nodes, up to 256 nodes (i.e., 2048 cores), using an
    8$\times$8$\times$4$\times$4 block size.  The right pane shows
    different convergence rates of SAP-FGCR as the Krylov subspace
    dimension is changed (32, 70, and 100), and the convergence
    achieved with SAP-FGMRES-DR with 10 deflated vectors and maximum
    Krylov subspace dimension of 70 (including the deflated vectors).
  }
\end{figure}

In Fig.~\ref{fig:sapdr} (left pane) we
show the scaling up to 256 nodes of the SAP preconditioner using an
8$\times$4$\times$8$\times$4 block size, compared with the ideal
scaling slope based on the performance measured on a single node.
We observe an excellent scaling behavior.

The condition number of the matrix $M$ depends on the simulation
parameters and can become very large.  The convergence of SAP-FGCR and
SAP-FGMRES then tends to become highly sensitive to the dimension of the
Krylov subspace.  This can be intuitively understood by noticing that
the solution of the Dirac equation in this regime is rich in low modes
that are not efficiently inverted by the SAP preconditioner. The outer
FGMRES iteration then has to reduce the norm of the residue associated
with those modes.

In order to improve the convergence properties of FGMRES we implemented
deflation in the form of \emph{deflated restart} (FGMRES-DR)
\cite{morgan}. This algorithmic trick allows one to preserve some of the
information about the Krylov subspace when restarting by reinserting
approximate eigenvectors into the Krylov subspace between restarts.
We note that in these cases the performance of the
algorithm severely depends on the dimension of the Krylov subspace,
which is limited by the amount of available memory.
This dependence is softer in case of FGMRES-DR.
In Fig.~\ref{fig:sapdr} (right pane) the convergence rate for solving
the same problem is compared using the two algorithms and different
Krylov subspace dimensions.
It would be interesting, but beyond the scope of the present paper, to
study the isoefficiency \cite{Kumar:1987} of our application taking
into account both algorithmic and machine performance.  From
Fig.~\ref{fig:sapdr} (right pane) we can deduce a (somewhat
counter-intuitive) example where increasing the number of nodes at
fixed problem size can increase the algorithmic efficiency since more
memory becomes available for the Krylov subspace.



\section{Summary and conclusions}\label{section:summary} 

In this paper we presented the strategies we applied to achieve
high-performance implementations of the most relevant kernel routines
on QPACE and to increase the fraction of code which leverages the compute
power of the accelerator cores of the {\cell} processor.

In an exploratory study it has been successfully shown that it is possible
to port a library which implements an application-specific software
interface to the Cell processor. This enables application programmers to
use the accelerator cores while the hardware details are completely hidden.
This goal has been accomplished by using advanced template programming
techniques and a code generator which generates code for the SPEs. On
top of this software interface the Chroma application suite has been
successfully built and tested.

When porting BQCD, an application for generating gauge configurations,
to QPACE we focused on an efficient solver for a linear set of equations.
This is the computationally most demanding part of the program.
A high machine performance has been obtained when using SAP-based
solvers. These algorithms allow us to perform a large number of operations
on the data stored in the on-chip memory, thus minimizing the amount of
data that is
moved in or out of the processor. Furthermore, large latencies for
node-to-node communications can be hidden.

The number of iterations needed by the SAP-based solvers strongly depends on
the dimension of the Krylov subspace, which is limited by the main memory
size. While hardware parameters typically impact the machine performance,
this is an interesting example where the algorithmic performance is
affected. This needs to be taken into account when defining optimal
machine parameters for a given algorithm.

In the summer of 2009 eight QPACE racks with an aggregate peak performance of
200 TFlops (double precision) have been deployed and are now used for
scientific applications, mainly from LQCD. The set of applications
running on QPACE also includes a fluid dynamics code based on the
discretized Boltzmann approach \cite{lbe}.

In the future we plan to apply the strategies we pursued for using the
accelerator cores of the {\cell} processor also to other architectures
with accelerator devices. For GP-GPUs highly optimized implementations
of relevant LQCD kernels are already available which, however, still
lack scalability to a large number of nodes (see, e.g.,
\cite{Babich:2010mu}). The implementation of software interfaces as
described in this paper will be more challenging due to the non-uniform
memory and the relatively slow bus connecting host processor and
accelerator device.

\section*{Acknowledgments}

We thank all members of the QPACE team for their hard and creative
work that laid the groundwork for the studies reported on in this
paper.
For the implementation of the SAP-based solvers the availability of a
reference implementation by M.~L\"uscher at
\url{http://luscher.web.cern.ch/luscher/DD-HMC} is gratefully acknowledged.
We also acknowledge the funding of the QPACE project provided by
the Deutsche Forschungsgemeinschaft (DFG) in the framework of
SFB/TR-55 and by IBM. We furthermore thank the following companies who
contributed significantly to the project in financial and/or technical
terms: Axe Motors (Italy), Eurotech (Italy), IBM, Kn\"urr (Germany),
Xilinx (USA), and Zollner (Germany).  This work was supported in part
by the European Union (grants 238353/ITN STRONGnet and
227431/HadronPhysics2).

We dedicate this paper to the memory of Gottfried Goldrian, who played
a key role in the development of the QPACE architecture.  He passed
away in January 2011.

\raggedright
\bibliographystyle{elsarticle-num}
\bibliography{qpace-iccs}

\begin{thebibliography}{10}
\expandafter\ifx\csname url\endcsname\relax
  \def\url#1{\texttt{#1}}\fi
\expandafter\ifx\csname urlprefix\endcsname\relax\def\urlprefix{URL }\fi
\expandafter\ifx\csname href\endcsname\relax
  \def\href#1#2{#2} \def\path#1{#1}\fi

\bibitem{apenext}
{F. Belletti et al.}, {Computing for LQCD: apeNEXT}, {Computing in Science and
  Engineering} 8 (2006) 18--29.
\newblock \href {http://dx.doi.org/10.1109/MCSE.2006.4}
  {\path{doi:10.1109/MCSE.2006.4}}.

\bibitem{qcdoc}
{P. A. Boyle et al.}, {Overview of the QCDSP and QCDOC computers}, {IBM J. Res.
  Dev.} 49 (2005) 351--365.
\newblock \href {http://dx.doi.org/10.1147/rd.492.0351}
  {\path{doi:10.1147/rd.492.0351}}.

\bibitem{cbea}
{H. P. Hofstee, A. K. Nanda, eds.}, {Cell Broadband Engine Technology and
  Systems}, {IBM J. Res. Dev.} 51 (2007) 501.

\bibitem{Baier:2009yq}
{H. Baier et al.}, {QPACE: A QCD parallel computer based on Cell processors},
  PoS LAT2009 (2009) 001.
\newblock \href {http://arxiv.org/abs/0911.2174} {\path{arXiv:0911.2174}}.

\bibitem{qpace-enahpc}
{H. Baier et al.}, {QPACE: power-efficient parallel architecture based on IBM
  PowerXCell 8i}, Computer Science - R{\&}D 25 (2010) 149--154.
\newblock \href {http://dx.doi.org/10.1007/s00450-010-0122-4}
  {\path{doi:10.1007/s00450-010-0122-4}}.

\bibitem{Pivanti:2011hd}
M.~Pivanti, S.~F. Schifano, H.~Simma, {An FPGA-based Torus Communication
  Network}, PoS LAT2010 (2010) 038.
\newblock \href {http://arxiv.org/abs/1102.2346} {\path{arXiv:1102.2346}}.

\bibitem{Nakamura:2010qh}
Y.~Nakamura, H.~St\"uben, {BQCD - Berlin Quantum Chromodynamics program}, PoS
  LAT2010 (2010) 040.
\newblock \href {http://arxiv.org/abs/1011.0199} {\path{arXiv:1011.0199}}.

\bibitem{Edwards:2004sx}
R.~G. Edwards, B.~Joo, {The Chroma software system for lattice QCD},
  Nucl.Phys.Proc.Suppl. 140 (2005) 832.
\newblock \href {http://arxiv.org/abs/hep-lat/0409003}
  {\path{arXiv:hep-lat/0409003}}, \href
  {http://dx.doi.org/10.1016/j.nuclphysbps.2004.11.254}
  {\path{doi:10.1016/j.nuclphysbps.2004.11.254}}.

\bibitem{veldhuizen}
T.~L. Veldhuizen, D.~Gannon, {Active Libraries: Rethinking the roles of
  compilers and libraries}, in: {In Proceedings of the SIAM Workshop OO’98},
  SIAM Press, 1998.
\newblock \href {http://arxiv.org/abs/math.NA/9810022}
  {\path{arXiv:math.NA/9810022}}.

\bibitem{winter-phd}
{F. Winter}, {Investigation of Hadron Matter using Lattice QCD and
  Implementation of Lattice QCD Applications on Heterogeneous Multicore
  Acceleration Processors}, Ph.D. thesis, {Regensburg University} ({2011}).

\bibitem{Belletti:2007pp}
{F. Belletti et al.}, {QCD on the Cell Broadband Engine}, PoS LAT2007 (2007)
  039.
\newblock \href {http://arxiv.org/abs/0710.2442} {\path{arXiv:0710.2442}}.

\bibitem{andrea-phd}
{A. Nobile}, {Performance Analysis and Optimization of LQCD Kernels on the Cell
  BE Processor}, Ph.D. thesis, {University Milano-Bicocca} ({2008}).

\bibitem{Luscher:2003qa}
M.~L\"uscher, {Solution of the Dirac equation in lattice QCD using a domain
  decomposition method}, Comput.Phys.Commun. 156 (2004) 209--220.
\newblock \href {http://arxiv.org/abs/hep-lat/0310048}
  {\path{arXiv:hep-lat/0310048}}, \href
  {http://dx.doi.org/10.1016/S0010-4655(03)00486-7}
  {\path{doi:10.1016/S0010-4655(03)00486-7}}.

\bibitem{saad}
{Y. Saad}, {Iterative methods for sparse linear systems}, 2nd Edition, {SIAM},
  {2003}.

\bibitem{itref}
C.~B. Moler, {Iterative Refinement in Floating Point}, {J. ACM} 14 ({1967})
  316--321.
\newblock \href {http://dx.doi.org/10.1145/321386.321394}
  {\path{doi:10.1145/321386.321394}}.

\bibitem{morgan}
R.~B. Morgan, {GMRES with deflated restarting}, SIAM Journal on Scientific
  Computing 24 ({2002}) 20--37.
\newblock \href {http://dx.doi.org/10.1137/S1064827599364659}
  {\path{doi:10.1137/S1064827599364659}}.

\bibitem{Kumar:1987}
V.~Kumar, V.~N. Rao, {Parallel depth first search. Part II. Analysis},
  International Journal of Parallel Programming 16 (1987) 501--519.
\newblock \href {http://dx.doi.org/10.1007/BF01389001}
  {\path{doi:10.1007/BF01389001}}.

\bibitem{lbe}
L.~{Biferale}, F.~{Mantovani}, M.~{Sbragaglia}, A.~{Scagliarini}, F.~{Toschi},
  R.~{Tripiccione}, {High resolution numerical study of Rayleigh-Taylor
  turbulence using a thermal lattice Boltzmann scheme}, Physics of Fluids 22
  (2010) 115112.
\newblock \href {http://arxiv.org/abs/1009.5483} {\path{arXiv:1009.5483}},
  \href {http://dx.doi.org/10.1063/1.3517295} {\path{doi:10.1063/1.3517295}}.

\bibitem{Babich:2010mu}
R.~Babich, M.~A. Clark, B.~Joo, {Parallelizing the QUDA Library for Multi-GPU
  Calculations in Lattice Quantum Chromodynamics}, {Proceedings of
  Supercomputing 2010.\ }\href {http://arxiv.org/abs/1011.0024}
  {\path{arXiv:1011.0024}}, \href {http://dx.doi.org/10.1109/SC.2010.40}
  {\path{doi:10.1109/SC.2010.40}}.

\end{thebibliography}

\end{document}